\documentclass[showpacs,amsmath,amssymb,twocolumn,aps]{revtex4}

\usepackage{amsfonts}
\usepackage{amsmath}
\usepackage{amssymb}
\usepackage{graphicx}
\usepackage{dcolumn}
\usepackage{times}
\usepackage{color}

\begin{document}

\title{Oscillons and oscillating kinks in the Abelian-Higgs model}

\author{V. Achilleos$^{1}$}
\author{F.K. Diakonos$^{1}$}
\author{D.J. Frantzeskakis$^{1}$}
\author{G.C. Katsimiga$^{1}$}
\author{X.N. Maintas$^{1}$}
\author{E. Manousakis$^{1,2}$}
\author{C.E. Tsagkarakis$^{1}$}
\author{A. Tsapalis$^{3}$}
\affiliation{$^{1}$Department of Physics, University of Athens, GR-15771 Athens, Greece}
\affiliation{$^{2}$Department of Physics, Florida State University, Tallahassee,
Florida, USA}
\affiliation{$^{3}$Hellenic Naval Academy, Hatzikyriakou Avenue, Pireaus 185 39, Greece}
\date{\today}

\begin{abstract}
We study the classical dynamics of the Abelian Higgs model
employing an asymptotic multiscale expansion method, which
uses the ratio of the Higgs to the gauge field amplitudes as a small parameter.
We derive an effective
nonlinear Schr\"{o}dinger equation for the gauge field,
and a linear equation for the scalar field containing the gauge field as a nonlinear source.
This equation is used to
predict the existence of oscillons and oscillating kinks for certain regimes of the ratio of the
Higgs to the gauge field masses. Results of direct numerical simulations are found to be in very good agreement with
the analytical findings, and show that the oscillons are robust, while kinks are unstable.
It is also demonstrated that oscillons emerge spontaneously as a result of the onset of the modulational instability
of plane wave solutions of the model. Connections of the results with
the phenomenology of superconductors is discussed.
\end{abstract}
\maketitle

\section{Introduction}
The concepts and tools of classical field theory are useful for a large class of physical systems ranging from the electromagnetic
field to
Bose-Einstein condensates (BECs).
For instance,
in the context of the Landau-Ginzburg (LG)
theory of phase transitions,
the free-energy density as a functional of the order parameter
(the expectation value of the quantum field operator for a BEC)
maps to the same form as the energy density functional of a classical
field.

%
Classical solutions minimizing the energy density functional of
such models play an important role also in quantum field theory, as well as in high energy physics, and cosmology. Real time soliton solutions give rise to particle-like structures, such as magnetic monopoles, and extended structures, such as domain walls and cosmic strings, that have implications for the cosmology of the early universe \cite{earlyuniverse,amin,stama068,khlopov}	. With respect to their dynamical characteristics, the classical field solutions can be both static or travelling solitons \cite{rajaraman}, as well as oscillons (alias breathers). Oscillons, are long-lived configurations localized in space and periodic in time. Such localized structures can be found in many physical systems, ranging from granular media \cite{granular} to water waves \cite{water}. Additionally, oscillons have also been extensively studied in the context of field theory, and can be found both in scalar \cite{sinegordon,bogo,campbell,gleiser94,honda,salmi06,salmi08,fodor06,fodor08,gleiser08,segur,amin2},
and Abelian or non Abelian gauge theories  \cite{gleiser2,farhi05,graham,sfakia}.
 Atomic Bose-Einstein condensates \cite{bec} offer a setting
where a large variety of classical field solutions
(relevant to elementary gauge fields \cite{spiel1} or even to
cosmological black holes \cite{bholes}), can be
studied both theoretically
and experimentally \cite{emergent}. Thus, the search for classical field solutions and their phenomenological impact, nowadays appears to gain an increased attention.

A prototype system of such a classical field emerges when one restricts himself to a mean-field treatment of the Gor'kov - Eliashberg - Landau - Ginzburg approach of superconductivity \cite{Gorkov,GorkElia,GinzbLand,Abrikosov}. In this case, the order parameter is the expectation value of an electron-pair creation operators which, for the simple case of s-wave pairing, behaves as a complex field $\phi$. The emerging effective field theory is an Abelian-Higgs model, where the Higgs field is the pair condensate order parameter, and the Abelian field is the electromagnetic field. This
explains in a rather straightforward manner the Meissner effect (the fact that the magnetic field only penetrates the superconductor within a finite penetration depth) through the now called Higgs mechanism, namely: the appearance of the condensate spontaneously breaks the $U(1)$ symmetry, giving rise to a finite mass to the gauge field. The existence of this gap (or mass) also ensures the phenomenon of perfect conductivity, through the collective dynamics of the condensed Cooper pair electrons. There is a considerable amount of works in the literature related to the search of classical solutions in this model (Abelian-Higgs), mainly in the context of localized and/or topologically charged forms. The existence of such solutions depends strongly on the space dimensionality. In two spatial dimensions, Abelian-Higgs models have vortex solutions (e.g., the Nielsen - Olesen vortex \cite{nielsen}), which carry a non-vanishing topological charge, and play an important role in many aspects of contemporary physics (for example the Abrikosov vortices in Landau-Ginzburg theory of superconductivity \cite{gl}). When time-dependence is taken into account, oscillon solutions can be found both in $(1+1)-$ \cite{mclerran} and $(2+1)-$ \cite{gleiser2} dimensions. Recently, it was shown that oscillons can emerge dynamically from  vortex-antivortex annihilation \cite{gleiser2}, highlighting the fact that oscillon dynamics are intriguing, and may play a crucial role in many areas of physics featuring symmetry breaking, ranging from superfluids and superconductors to cosmology.

In the present work, we
analyze the classical dynamics of the Abelian-Higgs model in $(1+1)$-dimensions
by means of a multiscale
expansion method \cite{multiplescales}.
We restrict our analysis on the case of small fluctuations of the condensate about its vacuum expectation value (vev). This is realized by choosing the amplitude of the Higgs field to be an order of magnitude smaller than that of the gauge field. 
Employing the method of multiple scales (see also Ref.~\cite{emeis}), the original nonlinear coupled field equations are 
reduced to an effective nonlinear Schr\"{o}dinger (NLS) equation for the gauge field, and a linear equation for the Higgs field containing the gauge field as a source. These equations are analytically solved giving two types of localized solutions
for the gauge field, namely in the form of oscillons and oscillating kinks.
Subsequently, we numerically integrate the original equations of motion using, as initial conditions,
the analytically found solutions.
We find that the analytical predictions are in excellent agreement with the numerical findings
for a large range of values of the parameters involved.
We then discuss connections between the Abelian-Higgs model solutions
and the phenomenology of superconducting materials, describing the temporal fluctuations of the condensate in a
one-dimensional (1D) Josephson junction~\cite{paterno} due to the presence of oscillating (in time) magnetic and electric fields.

The paper is organized as follows: in section II, we introduce the general formalism of the Abelian-Higgs model and we derive the equations of motion for the gauge and the scalar
fields.
We
then use the method of multiple scales to reduce the equations of motion
to the NLS system, and
present the corresponding analytical solutions.
In section III we present results of direct numerical integration of the original equations of motion
to check the validity of our approximations and study the stability of our solutions.
Finally, in section IV, we present our concluding remarks and make connections with the context of superconductors.


\section{The model and its analytical consideration}

\subsection{Formulation and Equations of Motion}

The $U(1)$-Higgs field dynamics is described by the Lagrangian:
\begin{eqnarray}
{\mathcal{L}}=-\frac{1}{4} F_{\mu \nu} F^{ \mu \nu} &+& (D_\mu \Phi)^{*}(D^\mu \Phi)
- V(\Phi^{*} \Phi),
\label{eq:eq1}
\end{eqnarray}
where $F_{\mu \nu}$ is the $U(1)$ field strength tensor, $D_\mu=\partial_{\mu} +ie \tilde{A}_{\mu}$ is the covariant
derivative ($e$ is the coupling constant), and $V(\Phi^{*} \Phi)=\mu^2 \Phi^{*} \Phi + \lambda (\Phi^{*} \Phi)^2$ ($\lambda > 0$)
is the Higgs self-interaction potential (asterisk denotes complex conjugate). In the broken phase, $\mu^2 <0$,
the scalar field acquires a non vanishing vacuum expectation value (vev) and can be parametrized as:
$\Phi=\frac{1}{\sqrt{2}}(\tilde{\upsilon} +\tilde{H})e^{i\chi/\tilde{\upsilon}}$, where the real functions $\tilde{H}$ and $\chi$
denote the Higgs field and the Goldstone mode, respectively,
and $\tilde{\upsilon}^2=-\mu^2 / \lambda$ is the corresponding vev. Since the Lagrangian of Eq.~(\ref{eq:eq1}) is
invariant under the local gauge transformations
\begin{eqnarray}
\Phi &\rightarrow & \Phi e^{ie \alpha(\mathbf{x},t)}  \\
\tilde{A}_{\mu}&\rightarrow &\tilde{A}_{\mu} - \partial_{\mu} \alpha(\mathbf{x},t),
\qquad \mathbf{x} \equiv (x,y,z),
\label{transf}
\end{eqnarray}
where $\alpha(\mathbf{x},t)$ is an arbitrary phase, by choosing the unitary gauge $\alpha=-\chi/\tilde{\upsilon}$, we can gauge away the Goldstone mode $\chi$.
This gauge choice has the advantage of mapping the $\Phi$ field to a real scalar field, namely the Higgs field.
In this case,
the fluctuations around the vev lead to the following system of equations of motion
for $\tilde{A}_{\mu}$ and $\tilde{H}$:
%
\begin{eqnarray}
(\widetilde{\Box} &+& m_A^2) \tilde{A}_{\mu} - \partial_{\mu} (\partial_{\nu} \tilde{A}^{\nu}) + 2e^2 \tilde{\upsilon}  \tilde{H} \tilde{A}_{\mu} + {e^2}  \tilde{H}^2 \tilde{A}_{\mu} = 0,
\nonumber \\ \label{eq:eq2}\\
(\widetilde{\Box} &+& m_H^2 )\tilde{H} + 3 \lambda \tilde{\upsilon} \tilde{H}^2 + \lambda \tilde{H}^3 - {e^2} \tilde{A}_{\mu} \tilde{A}^{\mu} (\tilde{\upsilon} + \tilde{H}) = 0,
\nonumber \\
\label{eq:eq3}
\end{eqnarray}
%
where $m_A^2={e^2 \tilde{\upsilon}^2}$ and $m_H^2=2 \lambda \tilde{\upsilon}^2$.
We simplify Eqs.~(\ref{eq:eq2})-(\ref{eq:eq3}) choosing the field representation $\tilde{A}_0=\tilde{A}_1=\tilde{A}_3=0$,
and $\tilde{A}_2=\tilde{A}\neq0$. Due to the fact that we are interested in 1D settings,
the non vanishing $\tilde{A}_2$ field depends solely on $x$ and $t$.
As a consequence, the Lorentz condition $\partial_{\nu} \tilde{A}^{\nu} = 0$ is fulfilled automatically,
leading to a coupled system of equations for the gauge field
$\tilde{A}(x,t)$ and the Higgs field $\tilde{H}(x,t)$.
Furthermore, we
express Eqs.~(\ref{eq:eq2})-(\ref{eq:eq3}) in a dimensionless form
by rescaling space-time coordinates and fields as: $x=x/m_A$, $t=t/m_A$, $\tilde{A}=(m_A/e) A$, $\tilde{H}=(m_A/e)H$.
This way, we end up with the following equations of motion:
\begin{eqnarray}
(\Box &+& 1)A + 2 A H + H^2 A = 0, \label{eq:eq4}\\
(\Box &+& q^2)H+\frac{1}{2}q^2 H^3 +\frac{3}{2}q^2 H^2 + H A^2 + A^2 = 0,
\label{eq:eq5}
\end{eqnarray}
where parameter $q \equiv m_H/m_A$ is assumed to be of order $\mathcal{O}(1)$.
The Hamiltonian corresponding to the above equations of motion is:
\begin{eqnarray}
\mathcal{H}= \frac{1}{2}\big[ (\partial_t A)^2 +  (\partial_x A)^2 +
(\partial_t H)^2 + (\partial_x H)^2 \big] + V,
\label{ham}
\end{eqnarray}
where indices denote partial derivatives
with respect to  $x$ and $t$, and the potential $V$ is given by:
\begin{eqnarray}
V = \frac{q^2}{8}H^4+\frac{q^2}{2} H^3 + \frac{q^2}{2} H^2 + \frac{1}{2} H^2  A^2 + H A^2 + \frac{1}{2} A^2. \nonumber \\
\label{pot}
\end{eqnarray}
Notice that $V$ exhibits a single minimum at $(A,H)=(0,0)$.

Here we should note that the above choice of field representation, and the concomitant simplification of the
full system into Eqs.~(\ref{eq:eq4})-(\ref{eq:eq5}), are valid in the case of the unitary gauge.
In a different gauge, the simplification would still be approximately valid, as long
as the Goldstone fluctuations $\delta\chi/\tilde{\upsilon}=\alpha+\chi/\tilde{\upsilon}$, are sufficiently smaller than the Higgs field $H$ or  slowly-varying with respect to the spatial variables.
However, if the Goldstone fluctuations become significant, Eqs.~(\ref{eq:eq4})-(\ref{eq:eq5})
should be modified to incorporate a phase field as well as more gauge field components.
In such a case, it is clear that a solution of Eqs.~(\ref{eq:eq4})-(\ref{eq:eq5}) (such as the ones that we will present below)
will not satisfy the full system. However, using continuation arguments, one can argue that for a sufficiently
small phase field and for additional gauge field components solutions of Eqs.~(\ref{eq:eq4})-(\ref{eq:eq5}) will still persist.
Further increase of the additional components may lead to bifurcations of these solutions and the emergence
of other structures. This problem is important and interesting in its own right, but its consideration is
beyond the scope of this work.

%
%
%
%
%
%
%

Thus, below we will focus on the simplified system (\ref{eq:eq4})-(\ref{eq:eq5}) and employ an asymptotic
multi-scale expansion method to derive approximate localized solutions, in the form of oscillons and oscillating kinks.


\subsection{
Multiscale expansion and the NLS equation}

Considering a small fluctuating field $H$,
we may employ a perturbation scheme, which uses a formal
small parameter $0<\epsilon\ll 1$
defined by the ratio of the amplitude of the Higgs field to the amplitude of the gauge field, i.e.,
$\epsilon \sim H/A$. In particular, we will employ a multiscale expansion method
\cite{multiplescales}, assuming that the Higgs and gauge fields depend on the set of independent variables
$x_0=x$, $x_1=\epsilon x$, $x_2=\epsilon^2 x, \ldots$ and
$t_0=t$, $t_1=\epsilon t$, $t_2=\epsilon^2 t, \ldots$.
Accordingly, the partial derivative operators
are given (via the chain rule) by $\partial_x = \partial_{x_0} + \epsilon \partial_{x_1} + \ldots$,
$\partial_t = \partial_{t_0} + \epsilon \partial_{t_1} + \ldots$.
Furthermore, taking into regard that the fields should be expanded around the trivial solution of
Eqs.~(\ref{eq:eq4})-(\ref{eq:eq5}), as well as $H \sim \epsilon A$ as per our assumption,
we introduce the following asymptotic expansions for $A$ and $H$:
\begin{eqnarray}
A&=& \epsilon A^{(1)} + \epsilon^2 A^{(2)} + \ldots, \label{A}
\\
H&=& \epsilon^2 H^{(2)} + \epsilon^3 H^{(3)}+ \ldots,  \label{eq:eq6}
\end{eqnarray}
where $A^{(j)}$ and $H^{(j)}$ ($j=0,1,\ldots$)
denote the fields at the order $\mathcal{O}(\epsilon^j)$.
Substituting Eqs.~(\ref{A})-(\ref{eq:eq6}) into
Eqs.~(\ref{eq:eq4})-(\ref{eq:eq5}), and using the variables $x_0$, $x_1,\ldots$, $t_0$, $t_1,\ldots$,
we obtain the following system of equations up to the third-order in the small parameter epsilon $\mathcal{O}(\epsilon^3)$:
\begin{eqnarray}
\!\!\!\!\!\!\!\!\!\!\!&\mathcal{O}(\epsilon)\!\!&:(\Box_0 + 1)A^{(1)}=0 ,
 \label{eq:order1}
 \\
\!\!\!\!\!\!\!\!\!\!\!&\mathcal{O}(\epsilon^2)\!\!& : \left(\Box_0 +1 \right)A^{(2)} + 2(\partial_{t_0} \partial_{t_1}-\partial_{x_0} \partial_{x_1})A^{(1)}=0, \label{eq:order2A}
\\
\!\!\!\!\!\!\!\!\!\!\!&\!\!&\;\;( \Box_0 + q^2)H^{(2)} + A^{(1)2}=0,
\label{eq:order2H}
\\
\!\!\!\!\!\!\!\!\!\!\!&\mathcal{O}(\epsilon^3)\!\!&:
\left(\Box_0 + 1\right)A^{(3)} + 2\left(\partial_{t_0}\partial_{t_1}-\partial_{x_0} \partial_{x_1}\right)A^{(2)}
\nonumber
\\
\!\!\!\!\!\!\!\!\!\!\!&\!\!& +\left(\Box_1 + 2\partial_{t_0} \partial_{t_2}-2\partial_{x_0} \partial_{x_2} + 2H^{(2)}\right)A^{(1)}=0.
 \label{eq:order3A}
\end{eqnarray}
The solution of Eq.~(\ref{eq:order1}) at the first order is:
\begin{equation}
A^{(1)}=u(x_1,x_2,\ldots,t_2,t_3\ldots)e^{i\theta} + {\rm c.c.},
\label{eq:eq10}
\end{equation}
where $u$ is an unknown function,
``c.c.'' stands for complex conjugate, while $\theta=Kx-\Omega t$, and the frequency $\Omega$
and wavenumber $K$ are connected
through the dispersion relation $\Omega^2=K^2+1$.
The solvability condition for the second-order equation~(\ref{eq:order2A}), is
$(\partial_{t_0} \partial_{t_1}-\partial_{x_0} \partial_{x_1})A^{(1)}=0$ (the second term is ``secular'',
i.e., is in resonance with the first term which becomes $\propto \theta \exp(i\theta)$, thus
leading to blow-up of the solution for $t \rightarrow +\infty$).
The above condition is satisfied in the frame moving with the group velocity $\upsilon_g$,
where $\upsilon_g \equiv \partial{\Omega}/\partial{K}$, so that $u$ depends only on the
variable $\tilde{x}_1=x_1-\upsilon_g t_1$ and $t_2$.
Then, the field $A^{(2)}$ is taken to be of the same functional form as $A^{(1)}$ [cf. Eq.~(\ref{eq:eq10})].
Taking into regard the above results, Eq.~(\ref{eq:order2H}) for the Higgs field yields:
\begin{equation}
H^{(2)}=B\left(b \vert u \vert^2 + u^2 e^{-2 it} +{\rm c.c.}
\right),
\label{eq:eq11}
\end{equation}
where $B = 1/(4-q^2)$ and $b=-2/Bq^2$; here, it is assumed that $q \ne 0$ and $q \ne 2$, as for
these limiting values the field $H^{(2)}$ (which is assumed to be of $\mathcal{O}(1)$ as per the perturbation expansion)
becomes infinitely large.
Finally, at the order $\mathcal{O}(\epsilon^3)$, we first note that the second term
vanishes automatically in the frame moving with $\upsilon$. Thus, the solvability condition
(obtained ---as before--- by requiring that the secular part vanishes) is:
$\left(\Box_1 + 2\partial_{t_0} \partial_{t_2} -2\partial_{x_0} \partial_{x_2}+ 2H^{(2)}\right)A^{(1)}=0$.
Below, for simplicity, we will consider the zero momentum case
($K=0$);
hence, $\upsilon_g=0$, and also $\tilde{x}_1=x_1$, i.e., the field $A_1$ is at rest.
In this case, we derive the following NLS equation
for the unknown function $u(x_1,t_2)$:
\begin{eqnarray}
i \partial_{t_2} u +\frac{1}{2} \partial_{x_1}^2 u  + s \vert u\vert^2 u = 0,\label{eq:eq12}
\end{eqnarray}
where the parameter $s$ is given by
\begin{eqnarray}
s &=& \frac{2}{q^2} + \frac{1}{q^2-4}. \label{s}
\end{eqnarray}

The above NLS equation possesses exact localized solutions of two different types,
depending on the sign of $s$: for $s>0$, the solutions are sech-shaped solitons, while for
$s<0$ the solutions are tanh-shaped kinks. These solutions have been extensively
studied in a variety of physical contexts including nonlinear optics \cite{yuri},
water waves \cite{johnson}, atomic Bose-Einstein condensates \cite{emergent}, and so on.
Here, the type of solution, i.e., the sign of the nonlinear term in Eq.~(\ref{eq:eq12}),
depends solely on the parameter $q$.

In particular, for $q<1.63$ and $q>2$, we obtain $s>0$ [attractive (focusing) nonlinearity] and the
soliton solution of Eq.~(\ref{eq:eq12}) has the following form:
\begin{equation}
u = u_0 {\rm sech} \left(\sqrt{s} u_{0} x_1 \right) e^{-i\omega t_2}, \label{eq:eq13}
\end{equation}
where $u_0$ is a free parameter [considered to be of order $\mathcal{O}(1)$] characterizing
the amplitude of the soliton, while the soliton frequency is $\omega = -(1/2) u^2_{0}s$.
Accordingly, the approximate solutions of Eqs.~(\ref{eq:eq4})-(\ref{eq:eq5}) for $A$ and $H$ can be expressed
in terms of the original variables $x$ and $t$ as follows:
\begin{eqnarray}
\!\!\!\!
A &\approx& 2\epsilon u_{0} {\rm sech}\left(\epsilon u_{0} \sqrt{s} x \right)
\cos\left[\left(1-\frac{1}{2}(\epsilon u_0)^2 s\right)t \right],
\label{eq:eq14} \\
\!\!\!\!
H &\approx & (\epsilon u_0)^2 B{\rm sech}^2\left(\epsilon u_{0} \sqrt{s}  x \right)  \phantom{aaaaaaaaaaaaaaaaa} \nonumber \\
&\times &\left\{ b+2\cos \left[2\left(1-\frac{1}{2}(\epsilon u_0)^2 s \right) t \right]\right\},
\label{eq:eq15}
\end{eqnarray}
and are characterized by a single free parameter, $\epsilon u_{0}$.
The above approximate solutions for the gauge field $A$ and the
Higgs field $H$ are localized in space (decaying for $|x|\rightarrow \infty$) and are oscillating in time; thus, they correspond to
{\it oscillons} (alias breathers). At the leading-order in $\epsilon$, the oscillation frequencies for $A$ and $H$ are respectively given by:
$\omega_A\approx 1$ and $\omega_H\approx 2$ (in units of $m_A$) respectively.

On the other hand, for $1.63<q<2$, we obtain $s<0$ [repulsive (defocusing) nonlinearity],
and Eq.~(\ref{eq:eq12}) possesses
a kink solution of the following form:
\begin{equation}
u= u_0 \tanh \left( \sqrt{|s|}u_{0} x_1 \right) e^{-i\omega t_2}, \label{eq:eq16}
\end{equation}
where $u_0$ is a $\mathcal{O}(1)$ free parameter and $\omega=u^2_0 |s|$. In this case, the approximate solutions
for the gauge and Higgs fields read:
\begin{eqnarray}
\!\!\!\!\!\!
A &\approx& 2 \epsilon u_{0} \tanh \left(\sqrt{|s|}\epsilon u_0 x \right)
\cos\left[\left(1+(\epsilon u_{0})^2 |s|\right)t\right], \label{eq:eq17} \\
\!\!\!\!\!\!
H &\approx& (\epsilon u_{0})^2 B \tanh^2 \left(\sqrt{|s|} \epsilon u_{0} x \right)  \phantom{aaaaaaaaaaaaaa}  \nonumber \\
&\times& \left\{b+2\cos \left[2\left(1+ \epsilon^2 u^2_{0}|s|\right)t\right]\right\}.
 \label{eq:eq18}
\end{eqnarray}
These solutions correspond to oscillating kinks, for both fields, which are zero exactly at the core of the kink
and acquire non vanishing values at $|x|\rightarrow \infty$. Again the fields oscillate with the frequencies
$\omega_A\approx 1$ and $\omega_H\approx 2$ respectively.

\section{Numerical results}
To further elaborate on our analytical findings, in this Section we will present results stemming from direct
numerical integration of Eqs.~(\ref{eq:eq4})-(\ref{eq:eq5}). Our aim is to check the validity of our analytical
predictions, namely the existence of oscillons and oscillating kinks, as well as study numerically their stability.
The equations of motion are integrated by using a fourth-order Runge-Kutta time integrator and
a pseudo-spectral method for the space derivatives \cite{jianke}. The lattice spacing used
was fixed to $\Delta x= 0.2$, the time step $\Delta t=10^{-2}$, and the total length of the lattice $L=400$.
In all simulations we used, as an initial condition, the approximate solutions of Eqs.~(\ref{eq:eq14})-(\ref{eq:eq15}) for the oscillons and of Eqs.~(\ref{eq:eq17})-(\ref{eq:eq18}) for the oscillating kinks for $t=0$;
in all cases we have fixed the free parameter value to $u_0=1$, and the ratio of the two fields amplitudes
to $\epsilon=0.1$.

\subsection{Oscillons}
\begin{figure}[tbp]
\includegraphics[scale=0.3]{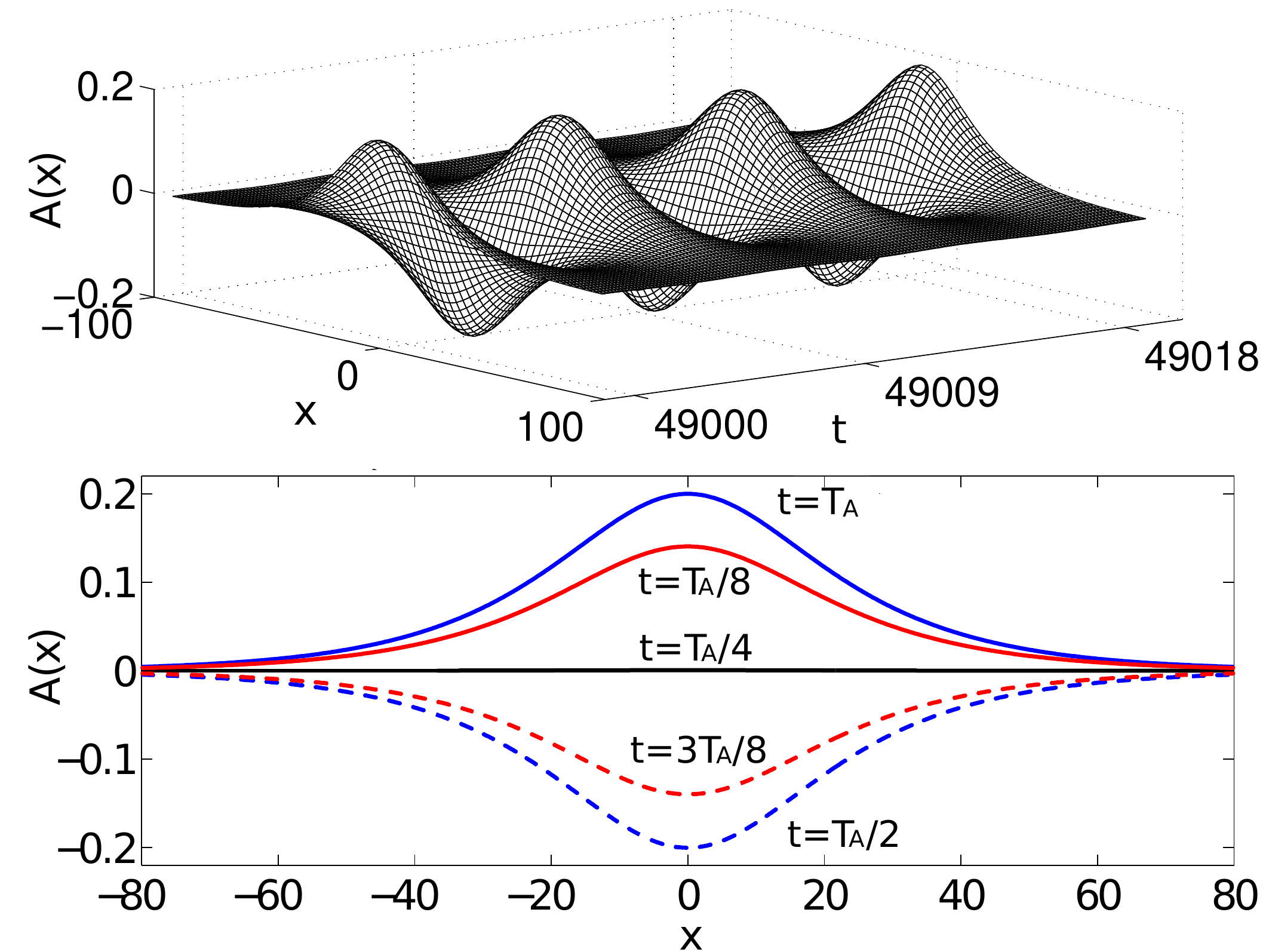}
\caption{(Color online) Top panel: Evolution of the gauge field $A$ as a function of $x$ and $t$. Bottom panel: Profile snapshots of the field at different instants of its breathing motion. Parameter values used are $\epsilon=0.1$, $q=1.5$ and the oscillation period is $T_A=2\pi$. }
\label{fig1}
\end{figure}
\begin{figure}[tbp]
\includegraphics[scale=0.3]{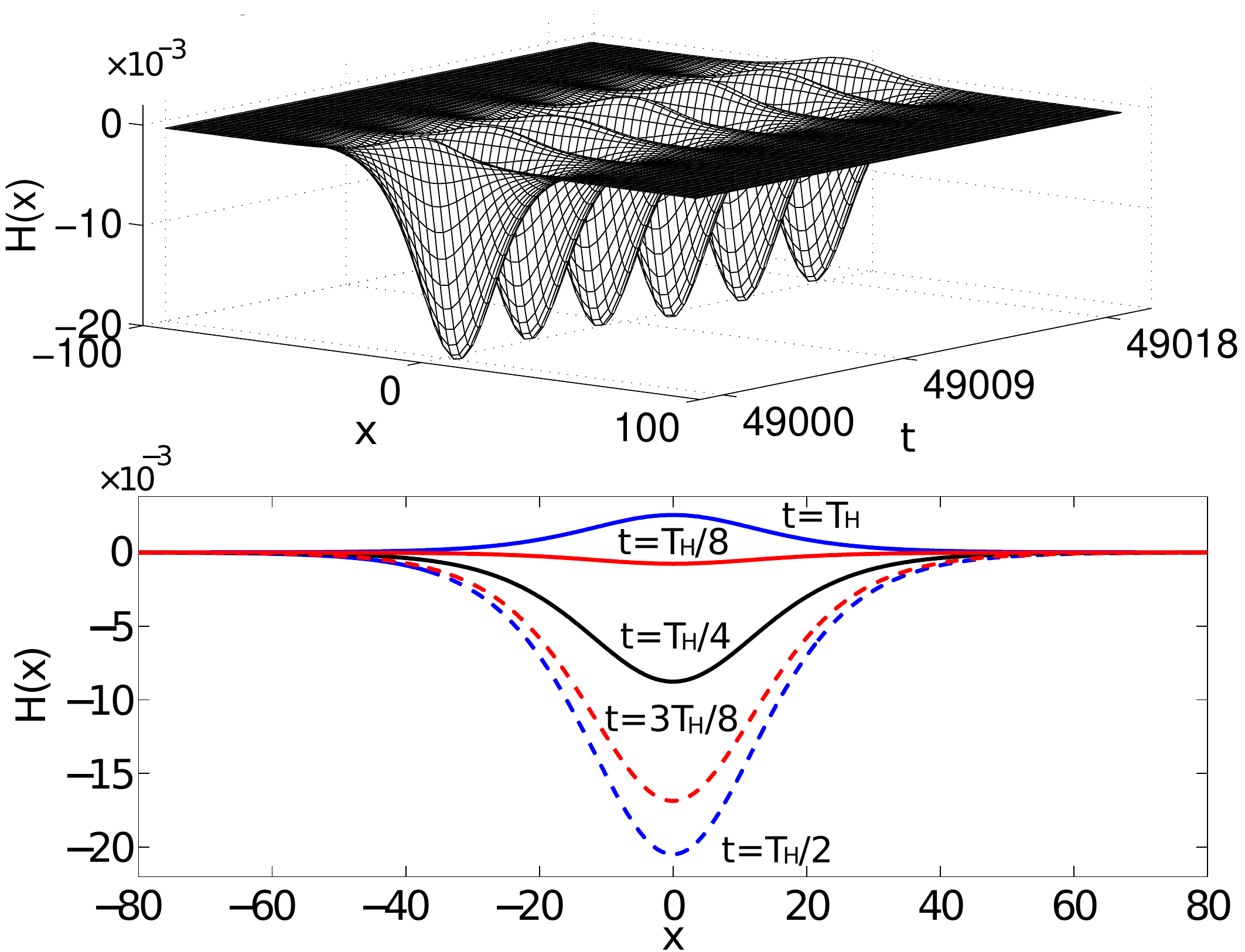}
\caption{(Color online) Same as in Fig.~\ref{fig1}, but for the
 Higgs field $H$. Notice that the oscilation period of the Higgs field is $T_H=\pi$.}
\label{fig2}
\end{figure}
\begin{figure}[tbp]
\includegraphics[scale=0.3]{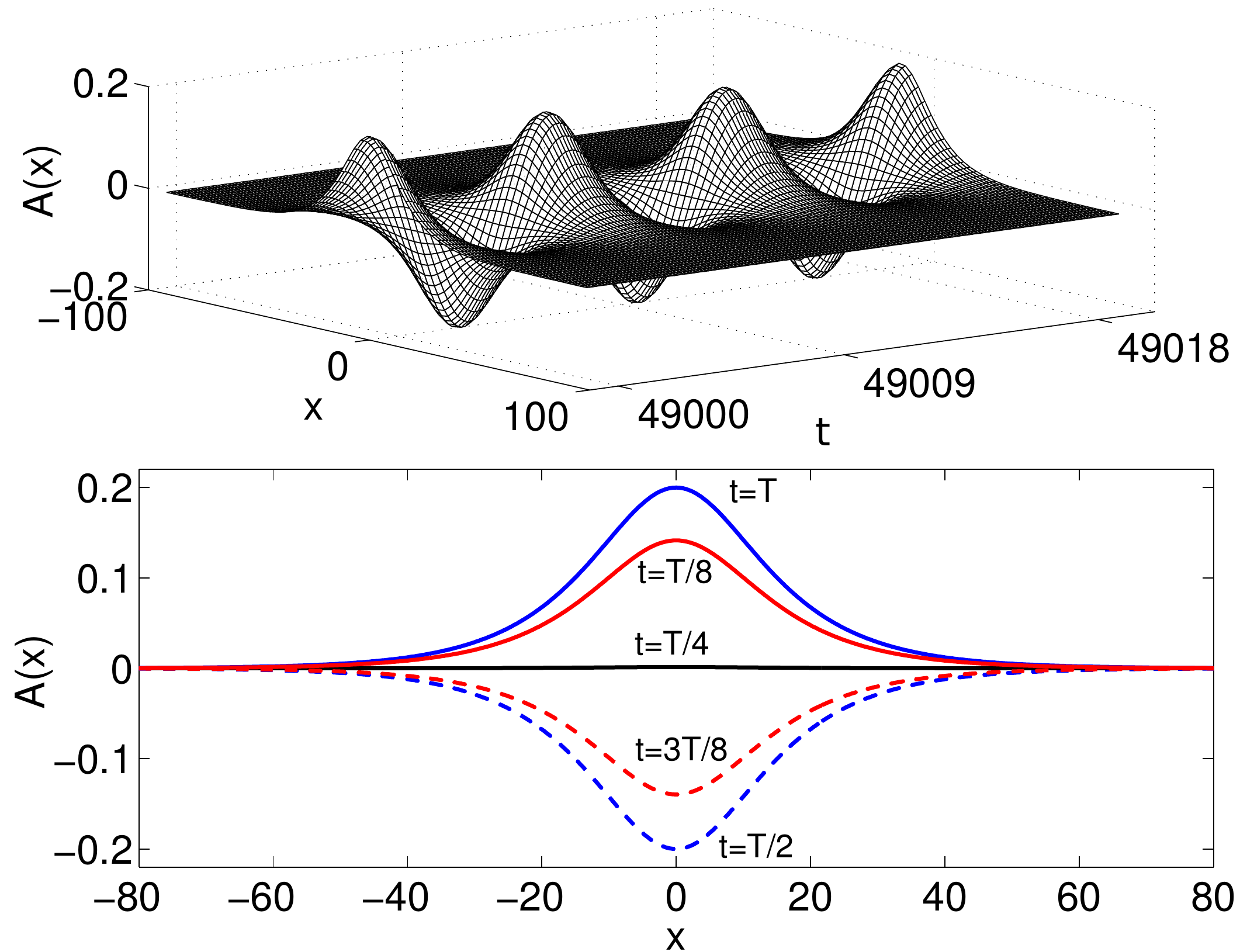}
\caption{(Color online) Same as Fig.~1, for parameter values $\epsilon=0.1$ and $q=2.5$. }
\label{fig3}
\end{figure}
\begin{figure}[t]
\includegraphics[scale=0.3]{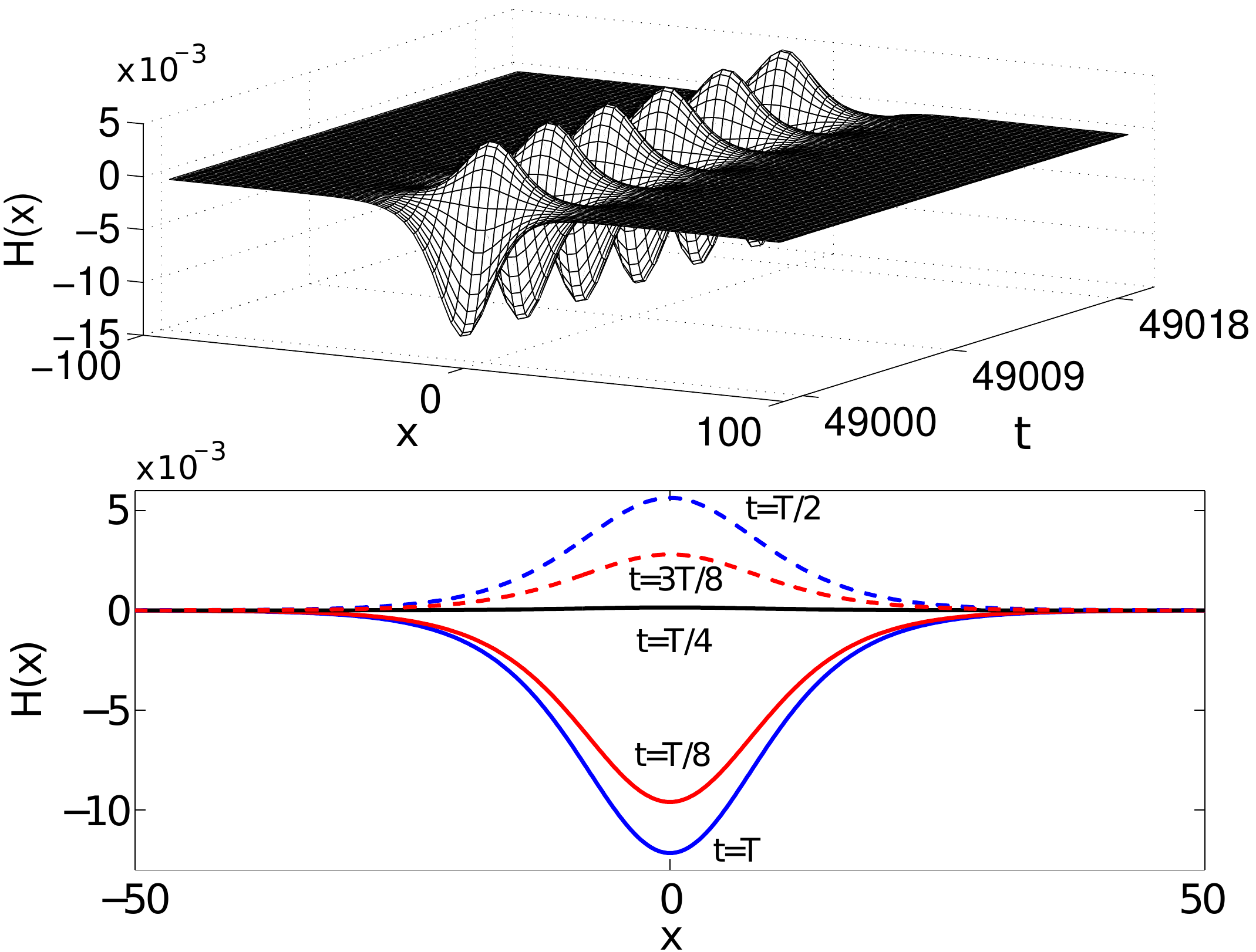}
\caption{(Color online) Same as Fig.~\ref{fig2}, with parameter values $\epsilon=0.1$, $q=2.5$. }
\label{fig4}
\end{figure}

In the top panels of Fig.~\ref{fig1} and Fig.~\ref{fig2} a contour plot showing the evolution of an
oscillon solution is shown (for $q=1.5$) at the end of the integration time
$t=5\times 10^{4}$. Note that the periods of
the two fields are $T_A\equiv 2\pi/\omega_A=2\pi$ and $T_H\equiv 2\pi/\omega_H=\pi$, and thus $3$ oscillations for the
gauge field and $6$ oscillations for the $H$ field are depicted.
In the bottom panels of Figs.~\ref{fig1} and \ref{fig2}, we show the profiles of $A$ and $H$ respectively,
during the interval of one period, exhibiting their breathing motion. While the gauge field $A$ performs symmetric
oscillations, the Higgs field oscillates asymmetrically with respect to its amplitude. This is due to the constant $b$
in the solution of Eq.~(\ref{eq:eq15}).
The evolution of an oscillon for
$q=2.5$ (lying in the second region where oscillons exist)
is also shown in the top panels of Fig.~\ref{fig3} and Fig.~\ref{fig4}, where due to the form of the solutions the
breathers become more localized in this region, as can be seen from their snapshots (bottom panels of
Figs.~\ref{fig3}-\ref{fig4}).

\begin{figure}[t]
\includegraphics[scale=0.4]{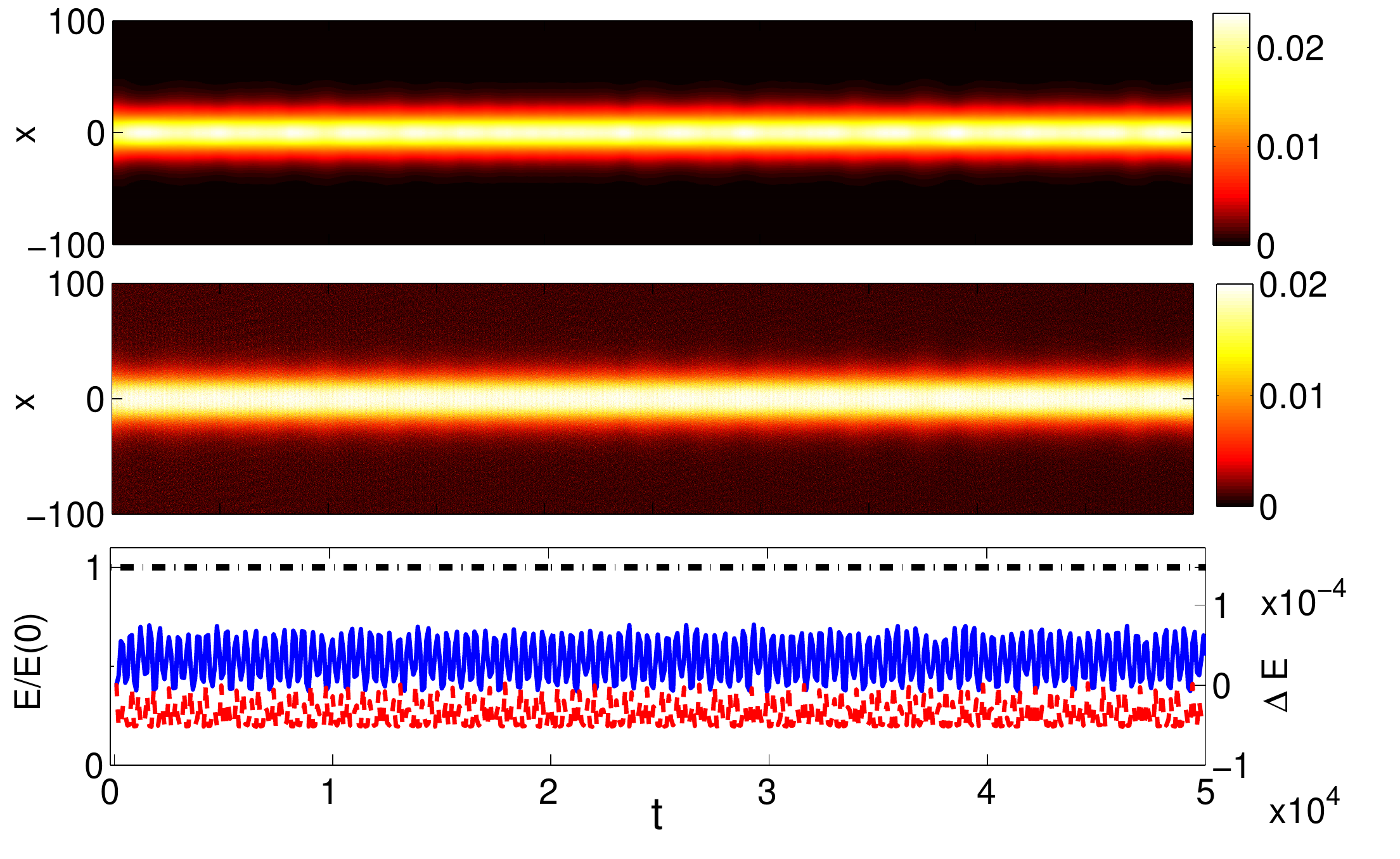}
\caption{(Color online) Top and middle panel: contour plot showing the energy density $\mathcal{H}(t)$, as
a function of time, for $q=1.5$.
For the middle panel the initial condition was perturbed by a small random noise with an amplitude $10\%$ of
the oscillon's amplitude. Bottom panel: the total energy normalized to its initial value at $t=0$,
where the dashed dotted [upper (black)] line corresponds to $q=1.5$ and $q=2.5$ respectively.
The energy difference $\Delta E$, for $q=1.5$  solid [middle (blue)] line and $q=2.5$ dashed [bottom (red)] line.}
\label{fig5}
\end{figure}

In order to also highlight the localization of the energy of the oscillons, in the top and middle panels of Fig.~\ref{fig5},
we show a contour of the total Hamiltonian density [cf. Eq.~(\ref{ham})] for $q=1.5$.
The top panel, corresponds to an initial condition using only Eqs.~(\ref{eq:eq14})-(\ref{eq:eq15}), while in the middle
panel we have also added a random Gaussian noise to the initial condition, as large as $10\%$ of the
oscillon's amplitude (the rest of the parameters are as in Fig.~\ref{fig1}).
We observe that the energy density remains localized during the numerical integration in both cases
(with and without noise).
We would like to stress here that, the latter result, shows that oscillons are robust
under the effect of a random noise, and this was also confirmed for different values of $q$ in
the domain of existence of the oscillon.
For completeness, the total energy $E(t)=\int^{\infty}_{-\infty} \mathcal{H}dx$, normalized to
its initial value $E(0)$, is also shown in the bottom panel of Fig.~\ref{fig5}; here,
the dashed dotted line corresponds to both $q=1.5$ and $q=2.5$. In fact the energy fluctuations,
defined as $\Delta E = E - E(0)$,
as shown in Fig.~\ref{fig5} [for $q=1.5$ (top solid line) and $q=2.5$ (bottom dashed line)]
are of the order of $10^{-4}$.


\subsection{Spontaneous oscillon formation}

In this section we show that oscillons can {\it emerge spontaneously}, through the mechanism of
modulation instability (for more details see Ref.~\cite{mi}). The
latter concerns the instability of
plane wave solutions of the NLS Eq.~(\ref{eq:eq12}),
of the form
\begin{equation}
u(x_1, t_2)= u_0 \exp{[i\left(kx_1 - \omega t_2\right)]}, \label{plane}
\end{equation}
under small perturbations.
To briefly describe the emergence of this instability, we consider
the following ansatz:
\begin{eqnarray}
u=(u_0+ W) \exp{[i\left(kx_1 - \omega t_2\right)+i\Theta)]},
\label{miansatz1}
\end{eqnarray}
where the amplitude and phase perturbations $W$ and $\Theta$ are given by:
\begin{eqnarray}
W&=& \varepsilon W_0 \exp{[i\left(Q x_1 - P t_2\right)]} + {\rm c.c.},
 \label{du}
 \\
\Theta &=&\varepsilon \Theta_0 \exp{[i\left(Q x_1 - P t_2\right)]}+ {\rm c.c.},
\label{dth}
\end{eqnarray}
with $W_0$ and $\Theta_0$ being constants,
and $\varepsilon$
being a formal small parameter.
Substituting Eqs.~(\ref{miansatz1})-(\ref{dth}) into Eq.~(\ref{eq:eq12})
it is found that, for $k=0$, the frequency $P$ and the wavenumber $Q$
of the perturbations, obey the dispersion relation:
\begin{eqnarray}
P^2=(1/2)|Q|^2(|Q|^2/2-2s|u_0|^2).
\label{dispmi}
\end{eqnarray}
It is evident from Eq.~(\ref{dispmi}) that for $s>0$ there exists an instability band
for wavenumbers $ Q^2< 4u_0^2 s$, where $P$ becomes complex. Note that this instability can only
emerge in the region where oscillons exist.
When this instability manifests itself, the exponential growth of the
perturbations leads to the generation of localized excitations, which are identified as the oscillons
described in Eqs.~(\ref{eq:eq14})-(\ref{eq:eq15}).

\begin{figure}[t]
\includegraphics[scale=0.4]{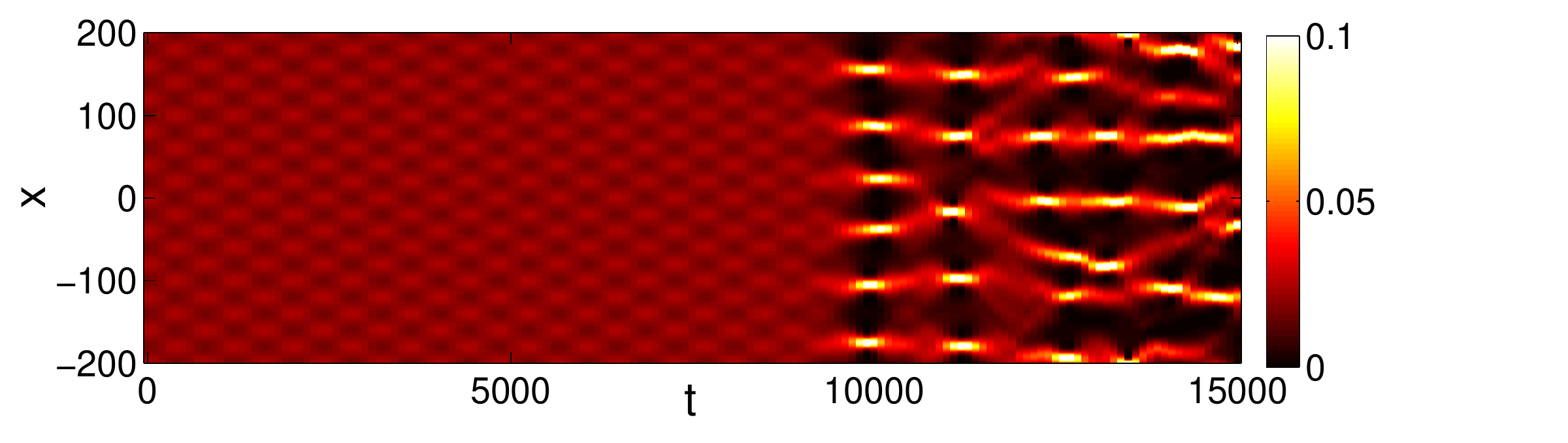}
\includegraphics[scale=0.4]{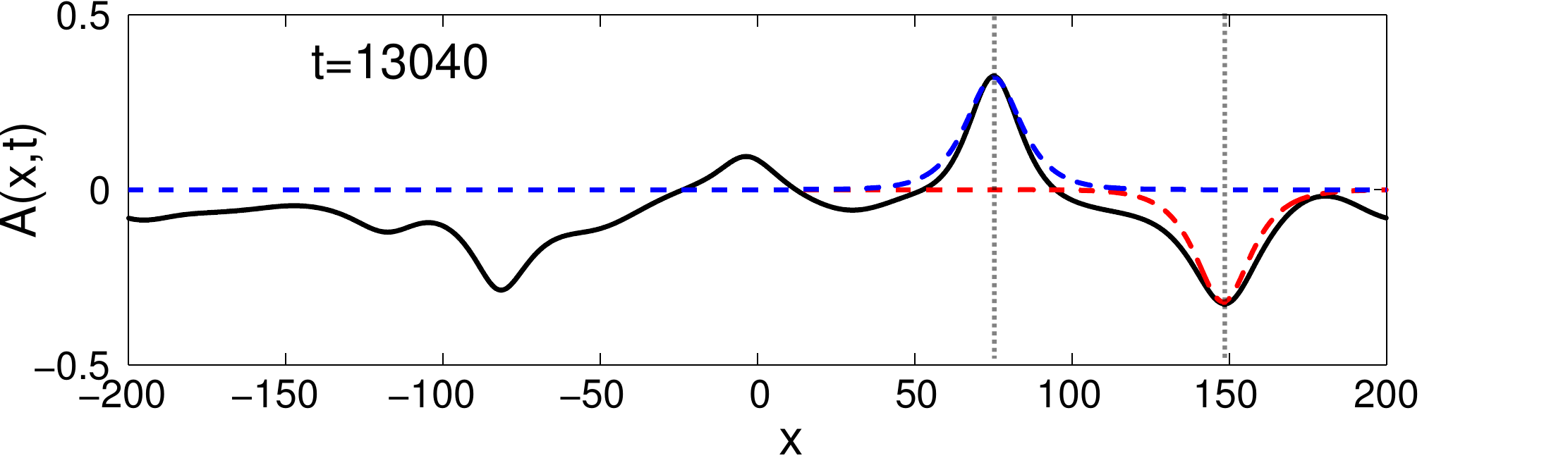}
\caption{(Color online) Top panel: Contour plot showing the energy functional $\mathcal{H}(t)$, for $q=1.5$, $Q=0.1$, $u_0=1$, $\epsilon=0.1$ for an initial condition corresponding to an unstable plane wave. Bottom panel: Profile of the gauge field $A(x)$, (solid [black] line), at $t=13040$ after the instability has set in. Dashed (blue and red) lines, correspond to the the solutions of Eq.~(\ref{eq:eq14}), fitted with the parameter $u_0$ at the peak of the oscillons (indicated by the dotted [grey] lines at $x=75$ and $x=150$). }
\label{fig7}
\end{figure}

To illustrate
the above, we have numerically integrated
Eqs.~(\ref{eq:eq4})-(\ref{eq:eq5}), with an initial condition corresponding to the plane wave of Eq.~(\ref{plane}),
perturbed as in Eq.~(\ref{miansatz1}), with $\varepsilon=0.1$, $W_0=1$, $Q=0.1$ (inside the instability band),
and $\Theta_0=0$; the rest of the parameters used are $q=1.5$, $ \epsilon = 0.1$, $u_0 = 1$ and $k=0$. In the top panel of Fig.~\ref{fig7},
we show a contour plot of the energy density, $\mathcal{H}(t)$. It is observed that, at $t\sim 10,000$, localization
of energy is observed due to the onset of the modulation instability. This localization is due to the
 fact that harmonics of the unstable wavenumber $Q$ of the perturbation are generated, which deform the
plane wave and lead to the formation of localized entities. In fact, the latter are eventually reformed
into oscillons, as is clearly observed in the
bottom panel of Fig.~\ref{fig7}: there, we show the profile of the gauge field $A(x)$ [solid (black) line]
at $t=13040$, and we identify at least two well formed oscillons located at $x=75$ and $x=150$.
The latter are found to be in a very good agreement with the respective analytical profile depicted by the dashed (red and blue) lines.
To perform the fitting, we plotted the approximate solutions of Eq.~(\ref{eq:eq14}), using the value at the peak of the oscillons,
 in order to identify the oscillon amplitude $u_0$. Then,
the solution was also displaced by a constant factor, in order to match the center of the oscillon. Additionally,
time was set to zero ($t=0$) and the fitting was made at the beginning of the period of oscillation for the particular oscillons. The
very good agreement between the analytical and numerical field profiles
highlights the fact that the oscillons considered in this section can be
generated spontaneously via the
modulational instability mechanism.      	

%
\subsection{Oscillating kinks}
\begin{figure}[t]
\includegraphics[scale=0.3]{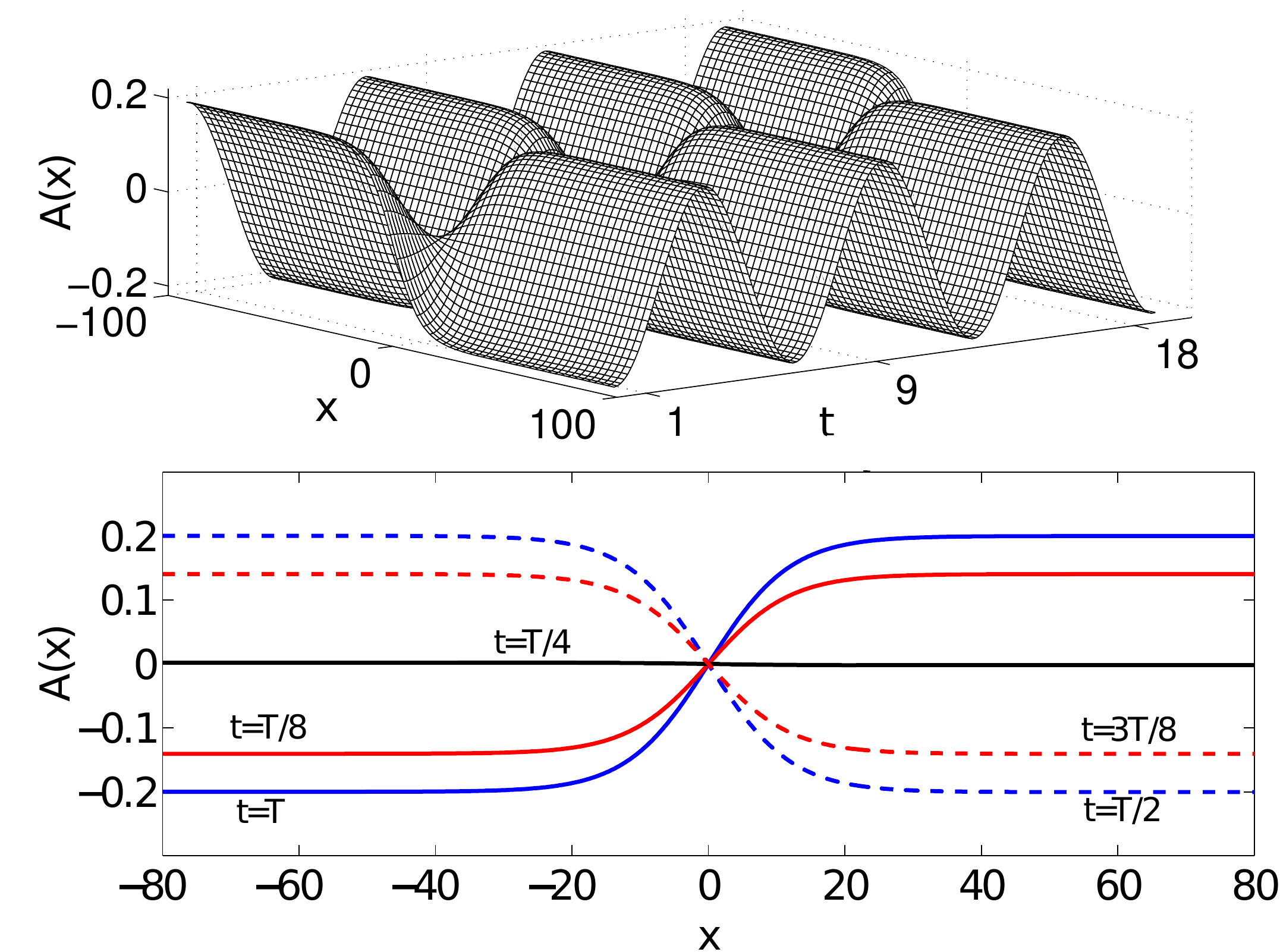}
\caption{(Color online) Same as Fig.~\ref{fig1}, but for a kink solution, for $q=1.8$.}
\label{fig8}
\end{figure}

\begin{figure}[t]
\includegraphics[scale=0.3]{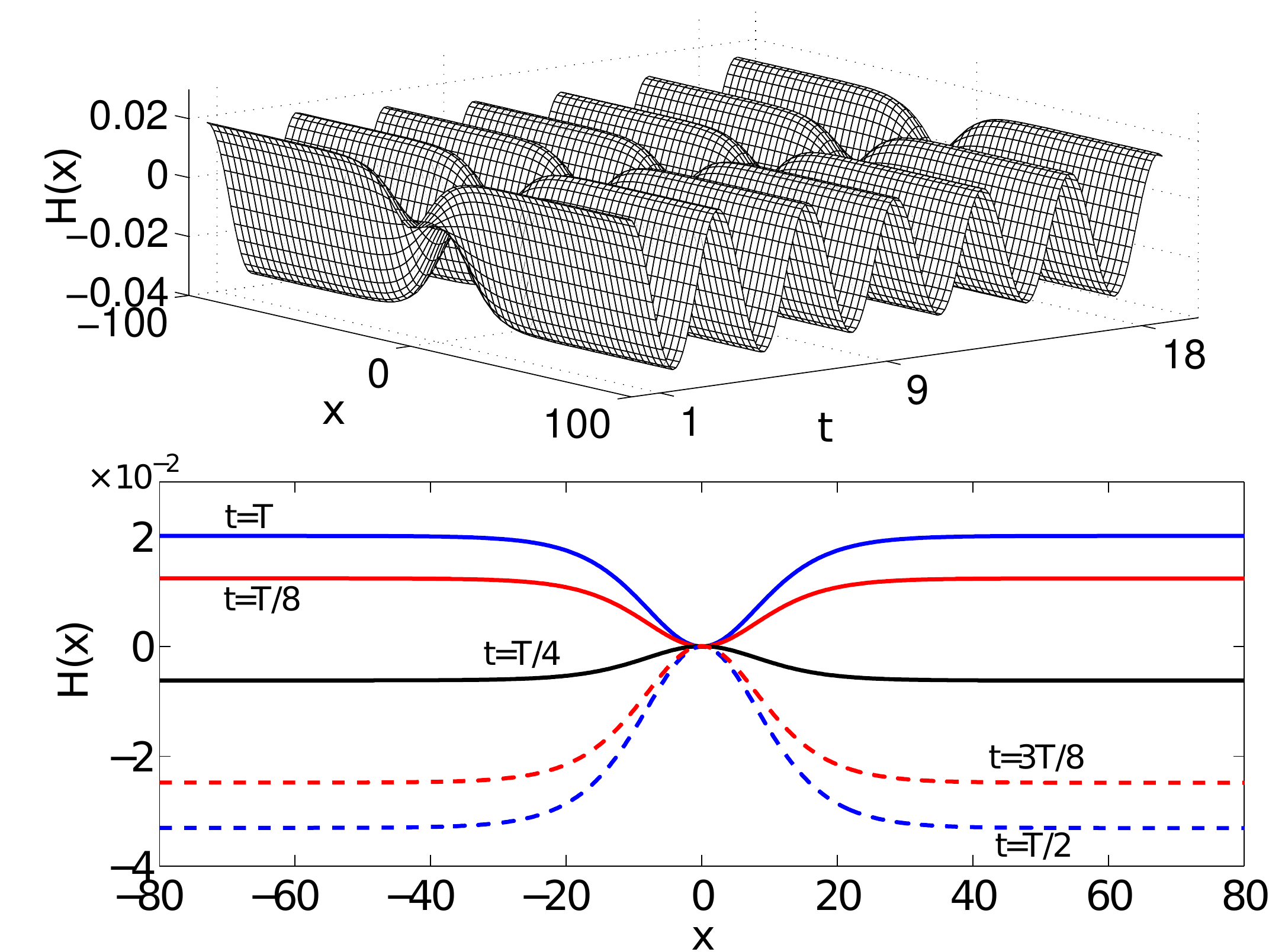}
\caption{(Color online) Same as Fig.~\ref{fig2} for the kink solution, and for $q=1.8$.}
\label{fig9}
\end{figure}

\begin{figure}[t]
\includegraphics[scale=0.3]{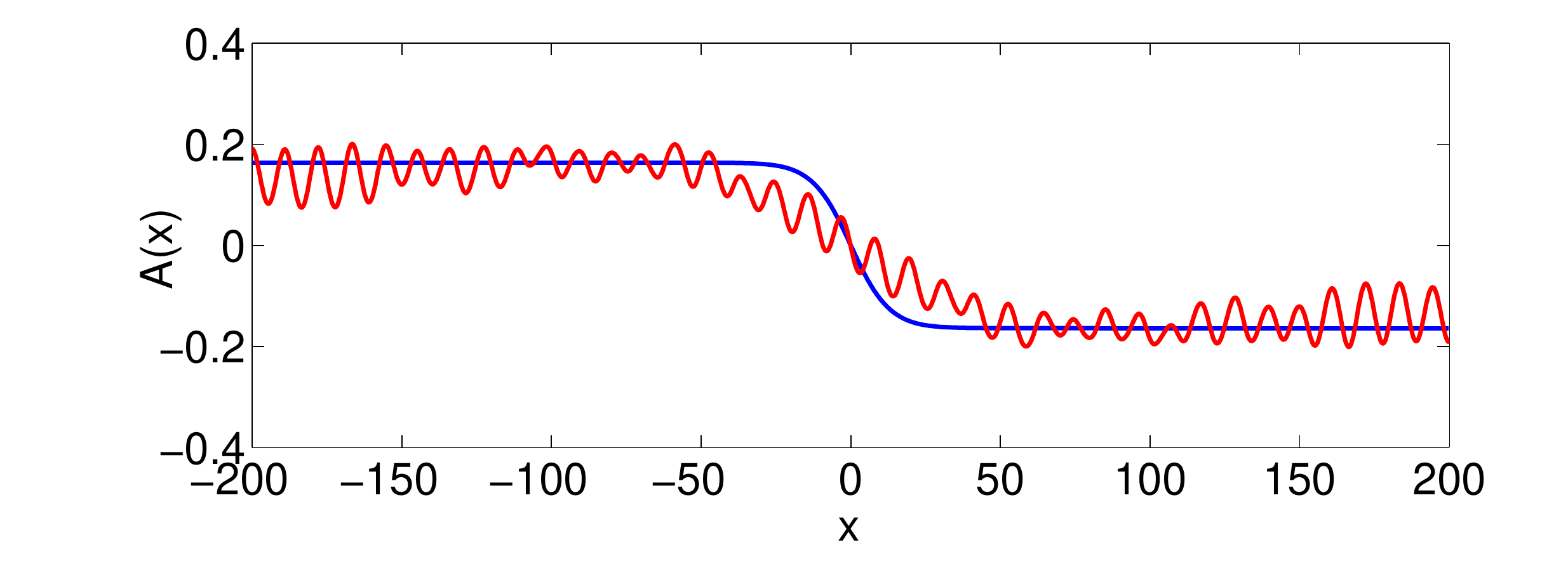}
\includegraphics[scale=0.3]{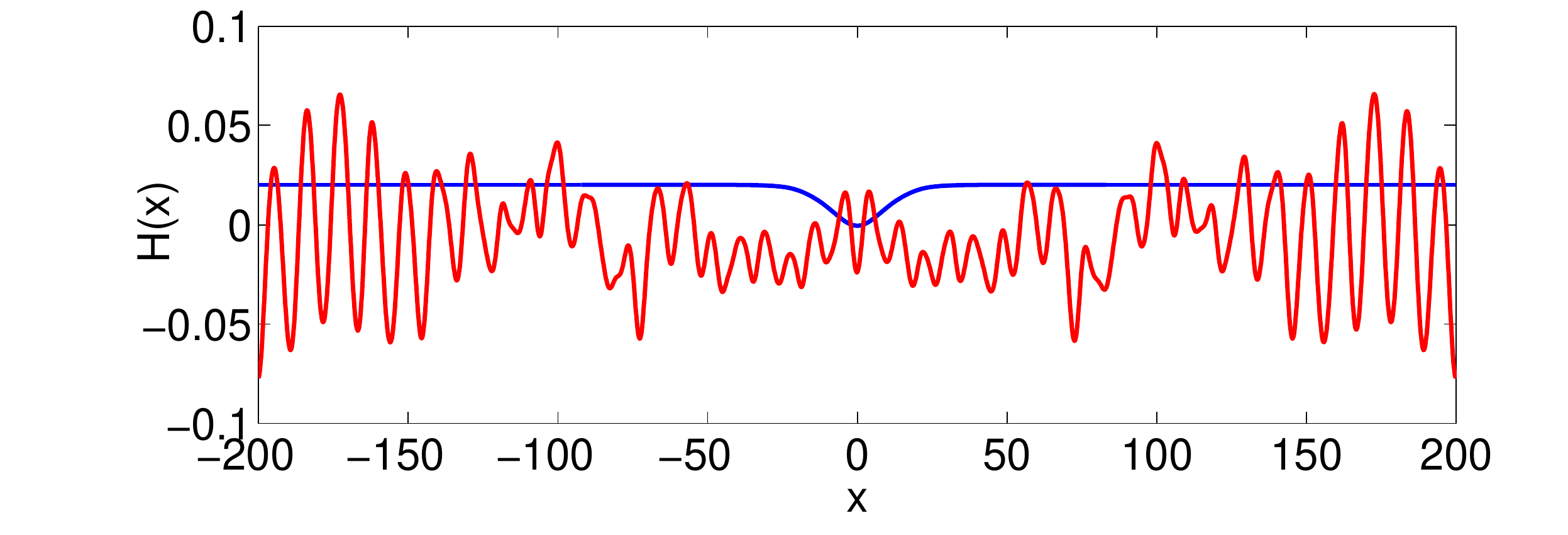}
\caption{(Color online) Top panel: profile snapshots of the gauge field at the beginning [solid (blue) line],
and at the end of the simulation [solid (red) line], for $q=1.8$. Bottom panel: same as in top panel but for the Higgs field.}
\label{fig10}
\end{figure}

We now
proceed with the numerical study of solutions corresponding to the kinks of the NLS Eq.~(\ref{eq:eq12}).
Such solutions in the form of Eqs.~(\ref{eq:eq17})-(\ref{eq:eq18}) are expected to
exist in the parameter region $1.63<q<2$. Performing the same procedure as in the previous section,
we numerically integrate the equations of motion
and study the relevant dynamics. In the top panels of Figs.~\ref{fig8} and \ref{fig9}, the contour plot
of a kink solution is plotted, showing the evolution of both fields $A$ and $H$, during the interval
of three and six periods, respectively (for $q=1.8$). In addition, the profiles of both fields are shown
in the bottom panels of Figs.~\ref{fig8}-~\ref{fig9}. The
oscillating kinks were found to be unstable for all values of the parameter $q$ within their region of existence. The
instability is manifested by an abrupt deformation of the kink solution, even near its core, characterized by the inverse
width $l_d$, (where $l_d= 2 \sqrt{|s|}\epsilon u_0$). In Fig.~\ref{fig10}, the top panel shows the profiles of the gauge
field at the beginning ($t=8$) [solid (blue) line], and at the end of the integration ($t=1200$) [solid (red) line].
While the shape of the kink is somehow preserved, it is clear that the solution profile has been significantly deformed.
More importantly, the Higgs field $H$, shown in the bottom panel, not only has been distorted but it has also
become an order of magnitude larger than its initial amplitude.
Thus, it can be concluded that oscillating kinks are unstable.


\begin{figure}[t]
\includegraphics[scale=0.3]{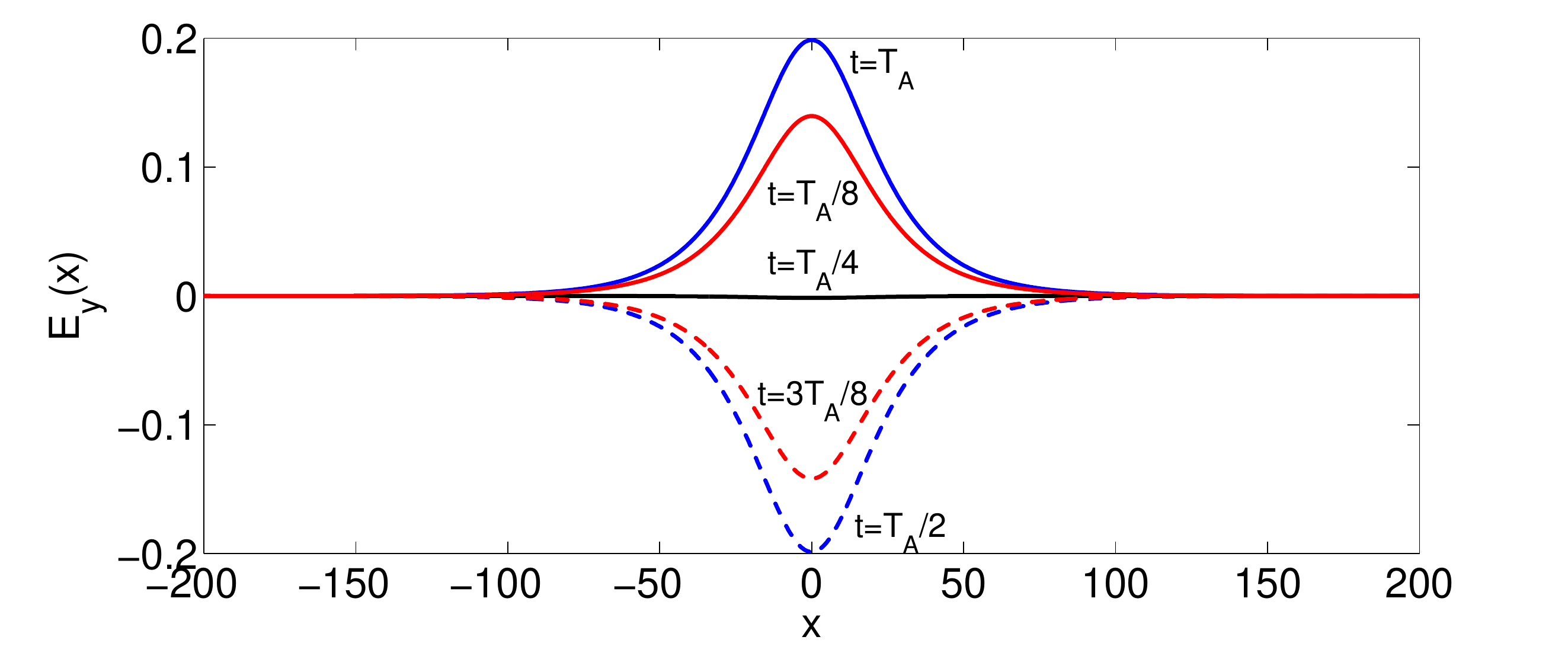}
\includegraphics[scale=0.3]{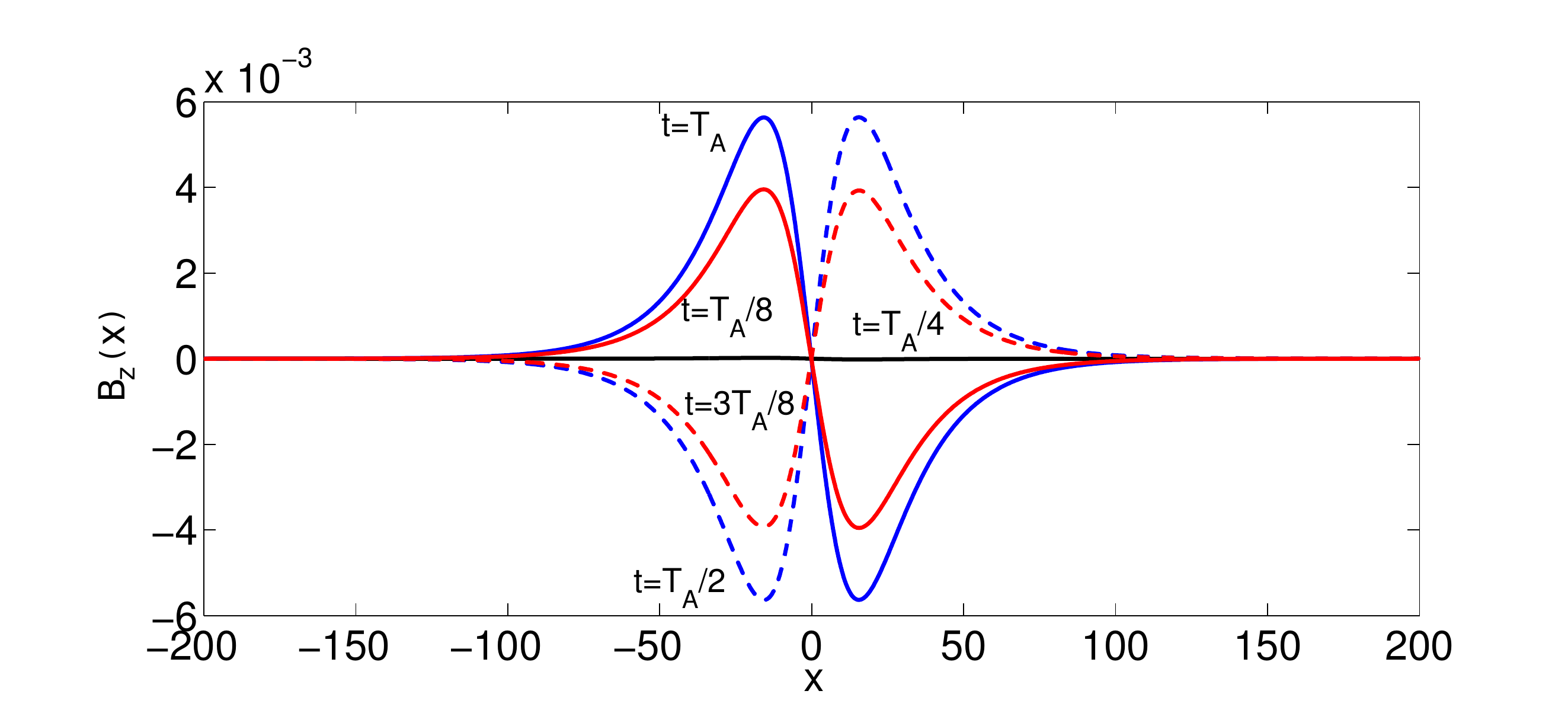}
\caption{(Color online) Top panel: profile snapshots of the electric field $\mathcal{E}_y (x)$ for $q=1.5$. Bottom panel: profile snapshots of the magnetic field $B_z (x)$ for the same $q$.}
\label{fig6}
\end{figure}
%


\section{Discussion and conclusions}

In this work we have presented a class of classical solutions of the 1D Abelian Higgs model obtained with the use
of a multiscale expansion method.
The key assumption in our treatment is that the scalar field amplitude is much weaker than that of the gauge field;
the relevant ratio was then used as a formal small parameter in the perturbation expansion.
We have shown that the equations of motion can be reduced to a NLS equation; by means of the latter,
localized solutions in the form of oscillons and oscillating kinks were derived.
Results obtained by direct numerical integration
of the original
equations of motion where found to be in very good agreement with the analytical findings.
In addition, we have numerically studied stability of the solutions against
a Gaussian noise with amplitude
up to $10\%$ of the Higgs field amplitude, and found that the oscillons remain robust.

It is also relevant
to discuss the possible connection of the
presented solutions to the physics of superconductors.
One could, in principle, write down the form of the magnetic and electric fields originating from the gauge field
$A$ which was chosen to be in the $\hat{y}$ direction:
$\vec{B}(x,t)=\partial_x A(x,t)\hat{z}$, and $
\vec{\mathcal{E}}(x,t)=-\partial_t A(x,t)\hat{y}$.
In the case of the oscillon solutions given in Eqs.~(\ref{eq:eq14})-(\ref{eq:eq15}), the above equations describe the electric field in he $y$ direction, which produces a magnetic field in the $z$ direction; both fields are localized around the origin of the $x$ axis -- cf.
Fig.~\ref{fig6}, where the profiles of the fields are shown.
This can be thought of as a configuration, describing
Superconductor--Normal metal--Superconductor (SNS)
Josephson junction~\cite{paterno}, where two superconductors are linked by a
thin normal conductor placed at the origin. Then, our solutions describe a condensate $\phi=(\upsilon+H(x,t))/\sqrt{2}$ that oscillates around its vev near the origin, and acquires its vev value when entering the superconductors. Accordingly,
the magnetic field is shown to oscillate inside the normal conductor, but vanishes exponentially inside the superconductors
as per the Meissner effect.

We would like to add that these oscillon solutions are different from the
phase-slip solutions which correspond to ``electric'' flux quantization
in close loops in one-time one-space dimensions. The latter solutions
correspond to an integer
number of electric flux quanta through a space-time loop \cite{ivlev,Langer}
and they are topological singularities which correspond to the change
of the phase of the condensate order parameter by a multiple of $2 \pi$ when
we ``travel'' around such a loop once. Oscillons on the other hand
are soliton solutions which are created by small-amplitude
excitations around the (superconducting) vacuum with the spontaneously
broken symmetry.

We note that it may be easier to observe such oscillon solutions more directly in rotating superfluids
and, in particular, in atomic Bose-Einstein condensates of trapped ultra-cold bosonic atoms
or in optical lattices in their superfluid phase.
The same U(1)-Higgs classical field theory may describe the dynamics of the condensate order parameter,
where the role of the external magnetic field is played by the local angular momentum of the rotating superfluid.

Our approach not only reveals a new class of solutions of the Abelian-Higgs model, but
also dictates a straightforward
general strategy for the search of non-trivial dynamics in models involving classical fields with
nonlinear interactions. It
is a natural
perspective to determine the impact of these solutions on the thermal and quantum behaviour of the
involved fields. Additionally, it would be extremely interesting to study the persistence and possible bifurcations
of the solutions presented in this work in the presence of the Goldstone mode and additional components of the gauge field.
In such a case, other nontrivial structures may emerge, whose stability and dynamics deserve to be investigated in detail.
Still another interesting direction is the consideration of higher-dimensional settings.
Such studies are currently in progress and relevant results will be reported elsewhere.

{\bf Acknowledgments.} Illuminating discussions with L. P. Gork'ov are kindly acknowledged.


\end{document}